\documentclass[twoside,a4paper]{article}
\usepackage[T2A]{fontenc}
\usepackage[cp1251]{inputenc}
\usepackage[english]{babel}
\usepackage[varg]{txfonts}
\usepackage{graphicx}
\makeatletter
\renewcommand{\@cite}[2]{{#1 \if@tempswa   #2\fi}}
\renewcommand{\@biblabel}[1]{\hfill}
\makeatother

\newcommand{\beq}{\begin{equation}}
\newcommand{\eeq}{\end{equation}}
\newcommand{\bea}{\begin{eqnarray}}
\newcommand{\eea}{\end{eqnarray}}

\def\sun{\hbox{$_\odot$}}

\begin{document}

\vbox{}

\begin{center}
{\LARGE  \bf Jets of SS\,433 on scales of dozens parsecs}
\vspace{0.7cm}

{\large  Alexander A. Panferov}

IMFIT, Togliatti State University, Russia

E-mail: panfS@yandex.ru
\end{center}
\vspace{0.7cm}

\begin{abstract}
The radio nebula W\,50 harbours the relativistic binary system SS\,433, which is
a source of the powerful wind and jets. The origin of W\,50 is wrapped in the 
interplay of the wind, supernova remnant and jets. The evolution of the jets on 
the scales of the nebula is a Rosetta stone for its origin.

To disentangle the roles of these components, we study physical conditions 
of the jets propagation inside W\,50, and determine deceleration of the jets.

The morphology and parameters of the interior of W\,50 are analyzed using the 
available observations of the eastern X-ray lobe, which traces the jet. 

In order to estimate deceleration of this jet, we devised a simplistic model of the viscous 
interaction, via turbulence, of a jet with the ambient medium, which would fit 
mass entrainment from the ambient medium into the jets of the 
radio galaxy 3C\,31, the well studied case of continuously decelerating jets.

X-ray observations suggest that the eastern jet persists through W\,50 as hollow 
one, and is recollimated to the opening $\sim 30^\circ$.
From the thermal emission of the eastern X-ray lobe, we determine a pressure of 
$P \sim 3\cdot 10^{-11}$\,erg/cm$^3$ inside W\,50. In the frame of a theory 
of the dynamics of radiative supernova remnants and stellar wind bubbles, this pressure in
combination with other known parameters restricts W\,50's origin to a supernova 
happened $\sim 100\,000$\,yr ago. Also, this pressure in our entrainment model 
gives a deceleration of the jet by $\sim 60\%$ in the bounds of W\,50's spherical 
component, of radius $\sim 40$\,pc. In this case, the age of the jet should be 
$\ll 27\,000$\,yr so as to satisfy the sphericity of W\,50.

The entrainment model comes to the viscous stress in a jet of a form 
$\sigma = \alpha P$, where the viscosity parameter $\alpha$ is predefined by the 
model.
\end{abstract}

\section{Introduction }
Astrophysical jets derive a significant part of energy released in accretion 
disks, and influence radically their environment. The powerful jets of SS\,433 
(\cite[1979]{AM79}) are so intimately interconnected with the surrounding 
shell-type radio nebula W\,50, that the nebula is thought to be not entirely of a 
supernova's origin. These jets are radiatively 
inefficient, that suggests their huge kinetic energy, of flux $L_\mathrm{k0} \sim
10^{39}$~erg/s, transforms into the thermal and mechanical energies of W\,50, 
and the latter is inflated partly by them. The revealed by \cite[(2015)]{Bor15}
gamma-ray emission, located within the W\,50 area, points one more item of the 
jets expenses: up to 10s of percentages of the jets energy might transform into 
relativistic particles. Moreover, it may indicate a region where the jets decelerate 
and transmit energy to the nebula: in the interior of W\,50, not at the nebula shell.

\begin{figure*}[ht]
\centering
\resizebox{\hsize}{!}{\includegraphics{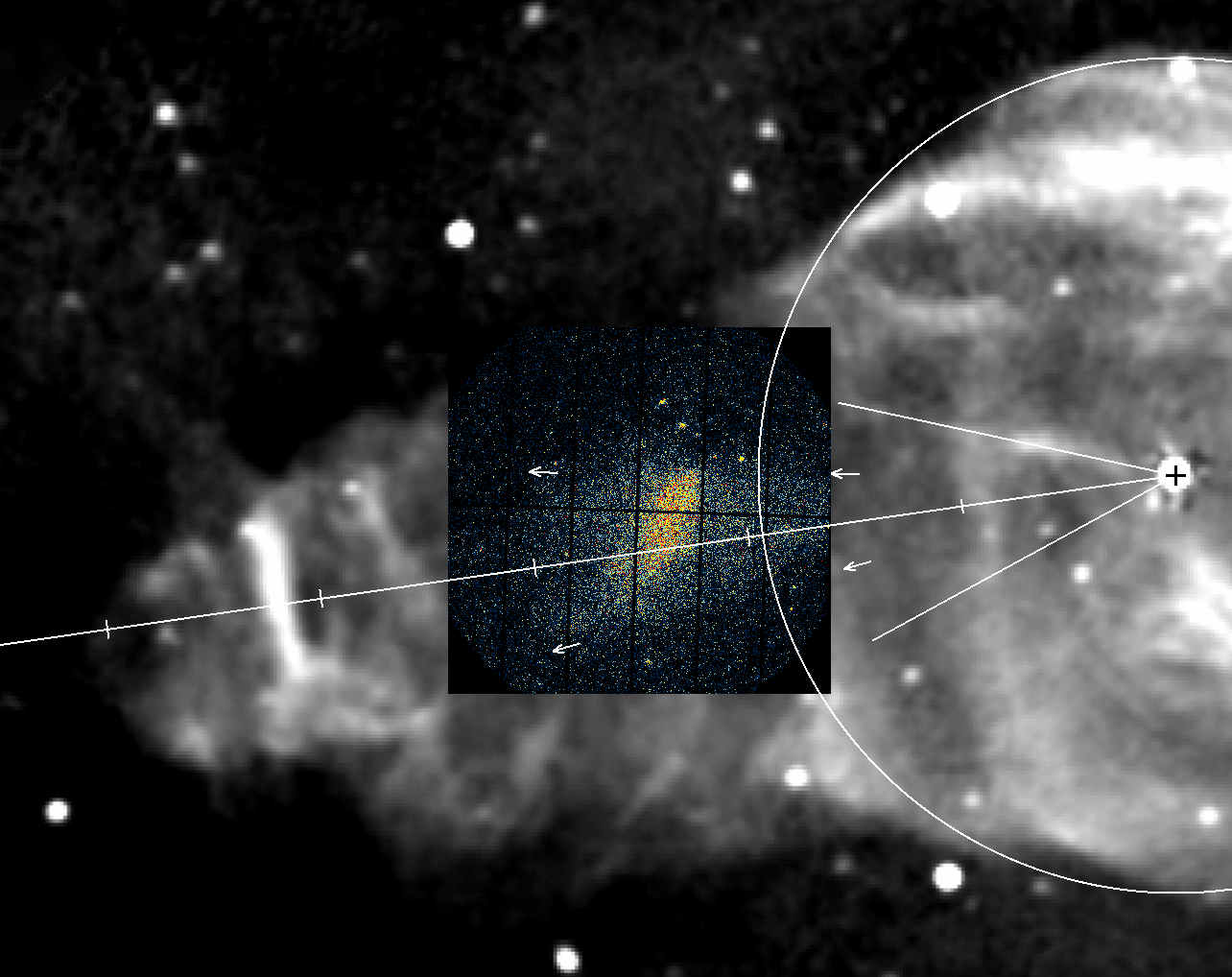}}
\caption{
The 1--2\,keV image of the bright knot region (the XMM-Newton observatory),
laid over W\,50's radio image (\cite[1998]{Dub98}), depicts the geometry of 
the eastern X-ray lobe in the hard emission. The images are given at the
epoch J2000, north is up, east is left, SS\,433 ($\mathrm{RA} = 
19^\mathrm{h}11^\mathrm{m}49.\!\!^\mathrm{s}57$, $\mathrm{Dec.} = 
04^\circ 58^\prime 57.\!\!^{\prime\prime}9$) is on the right. There are 
indicated the spherical component of W\,50, the precession cone and its axis, 
ticked every $15^\prime$ from the beginning at SS\,433. The borders of 
\cite['s\ (1983)]{Wat83} cut, for the radial profile of brightness, are 
delineated by arrows.
}
\label{Xcone}
\end{figure*}

The role of the jets of SS\,433 in formation of the peculiar W\,50 is unknown 
(see \cite[(2016)]{Far16} for a comprehensive review of the origin of W\,50).
The morphology of W\,50 in radio emission closely resembles an achatina, the 
giant African snail (\cite[1998]{Dub98}; see also Fig.~\ref{Xcone}). 
W\,50 looks like consisting of coils, narrowing to the tips of
the nebula like a pyramid. The torque and elongated appearance of W\,50 might
be due to the jets, which, however, are not explicitly resolved at the scales of
W\,50 ($1^\circ \times 2^\circ$). By a convention, one discerns in W\,50 a spherical
component, of radius $\sim 29^\prime$ (\cite[1998]{Dub98}), and two protrusions, 
the so called ears.

%******************************************************************
%
%
%******************************************************************

W\,50 is elongated along the jets, which are observed closely to SS\,433 as 
outflows in X-ray, optical and radio bands. At distances from the jets source 
up to $\sim 6''$, or at distances $r\lesssim 0.13$\,pc\footnote{$1'' = 
0.67\cdot 10^{17}$~cm at the observer distance of $D=4.5$\,kpc, accepted 
hereinafter (\cite[2002]{St02}; \cite[2010]{Pan10}; \cite[2013]{Mar13}; 
see also \cite[2007]{Loc07}, and \cite[2014]{Pan14} for discussion on the 
distance); data from the literature are compiled to this distance.}, the 
jets show a regular precession with period $162.250$\,days, under angle
$19.\!\!^{\circ}75$, around an axis whose inclination
to the line of sight is $78.\!\!^{\circ}8$ (\cite[2008]{Dav08}) and position 
angle on the plane of the sky is $98.\!\!^{\circ}2$ (\cite[2002]{St02}). At these 
distances the jets are ballistic, unchanging (e.g. Fig.~2 of \cite[2010]{Rob10}), 
except may be the predicted 10\% deceleration and twist (the shift of the 
precession phase by $\sim -0.1$) in the innermost $\sim 0.\!\!^{\prime\prime}5$ 
(\cite[2014]{Pan14}; see also \cite[2004]{St04}). At 
larger distances the jets become unobservable, possibly because of weakening
of the interaction of the jet with the ambient medium. The jet there would look like 
a hollow cone of tightly wound coils, probably blending. The cone of the 
precession is indicated on the radio image (\cite[1998]{Dub98}) of the eastern 
part of W\,50 in Fig.~\ref{Xcone} -- evidently, the cone fits the orientation and 
transverse size of the ear. The overlaid X-ray image\footnote{The HEASARC data 
archive, http://heasarc.gsfc.nasa.gov/FTP/xmm/ data/rev0//0075140401/PPS/,  
the EPIC PN camera of the XMM-Newton observatory.} in Fig.~\ref{Xcone}
suggests that the lobe of the X-ray emission, observed at distances 
$> 15^\prime$ from SS\,433 (\cite[1983]{Wat83}; \cite[1996]{Bri96}), shapes the 
jet. However, the angular extension of the lobe in the hard X-ray band, $>1$\,keV, 
is much smaller than the opening $40^{\circ}$ of the precession cone. Are the jets 
recollimated? On the contrary, the optical filaments at the contact between the 
ears and sphere of W\,50 (\cite[2007, Fig.~1 and 2]{Bou07}), which possibly 
trace the interaction of the jet and spherical shell of W\,50, subtend an angle 
a little more than $40^{\circ}$ at SS\,433. The eastern lobe, which is more exposed
in observations, has sharp edges in the hard X-ray band and is almost perfectly 
axisymmetrical and smooth, except the bright knot (Fig.~\ref{Xcone}; 
\cite[2007, Fig.~3 and 8]{Bri07}). The latter is probably the segment of the nearly 
spherical shell, heated by the jet (\cite[1983]{Wat83}). 

More than 30 years as the problem of the recollimation of the SS\,433 jets inspires
its investigation. \cite[(1983)]{Eic83} has proposed that a precessing jet merges in 
a smooth hollow cone and inevitably undergoes focusing by the ambient pressure. 
\cite[(1990)]{Koc90}
have ascertained the problems staying before the hydrodynamical simulations of 
Eichler's mechanism and of the pointed form of the W\,50 ears for the case of a 
hollow conical jet. Later, from a series of hydrodynamical simulations 
targeted at the ears geometry, \cite[(2011b)]{Go11b} have devised a history of 
the jet evolution intermittent in the speed and precession cone opening; 
moreover, they got the recollimation mechanism at work, although inefficient for 
the formation of the ears of the observed narrowness. However, there are not seen 
the traits of the intermittency, known for intermittent astrophysical jets: in W\,50 
the X-ray lobes are rather smooth at large scales. 

In particular, \cite[(2011b)]{Go11b} resume that in the case of a permanency
the SS\,433 jet should decelerate only at the terminal shock, i.e. in the ear. On
other hand, from the non-observation of the proper motion of the radio filaments in the
ears, \cite[(2011a)]{Go11a} received that the jets velocity in the ears is at least 
eight times smaller (for the distance $D=4.5$\,kpc) than the optical jets velocity 
$v_\mathrm{j0} = 0.2581c$, $c$ being the speed of light (\cite[2008]{Dav08}). 
Besides, the brightness of the X-ray ring-like structure at distances 
$\sim 60^\prime$, thought to be a terminal shock, coincident with the 
radio filaments in the eastern ear, is much 
smaller than one of the X-ray lobe at 15--$45^\prime$ from SS\,433 
(\cite[2007, Fig.~1 and 2]{Bri07}). It is hardly to explain unless the jets 
decelerate in the interior of W\,50. 

Thus, the questions about the evolution of the jets of SS\,433 at W\,50's scale 
and their role in the inflation of W\,50 remain. This paper solves the key 
question: Could the jets decelerate in the interior of W\,50, before the 
termination in the ears? Firstly, in Sect.~\ref{PhC} we characterize the 
physical conditions of the jets propagation in the eastern X-ray lobe using 
the available X-ray observations, namely the 
density and pressure in the surroundings of the jet. On the basis of the found
pressure, the age of the nebula W\,50 is evaluated, and the roles of a supernova 
remnant (SNR), SS\,433 system's wind, and the jets in the nebula formation are 
clarified in 
Sect.~\ref{Age}. Encouraged by the studies of the decelerating relativistic jets 
of the Fanaroff-Riley class I radio galaxies, in Sect.~\ref{Entr1} we construct
the model which would fit mass entrainment from the environment into the 
jets of the radio galaxy 3C\,31, the well known case of continuously decelerating 
jets. The entrainment model is applied to the 
SS\,433 jet, and results on the jet behaviour at scales of dozens parsecs
are presented in Sect.~\ref{SS}. The conclusions are given in Sect.~\ref{Iss}.

\section{\large Physical conditions inside the radio nebula W\,50}
\label{PhC}
Deceleration of the SS\,433 jets issues from the interaction of the jets with the 
ambient medium. Physical conditions of the jets propagation are little known. 
From the Einstein observatory's observations, \cite[(1983)]{Wat83} estimated 
temperature $kT \sim 2$\,keV and average electron number density $n_\mathrm{e}
\sim 0.1$\,cm$^{-3}$ between $15^\prime$ and $42^\prime$ from SS\,433, in 
the cone of half-angle $\sim 25^{\circ}$, with an apex at SS\,433, aligned with 
the eastern jet axis, and the density range $\sim 0.16$--0.7\,cm$^{-3}$ between the 
faintest, at $r>45^\prime$, and brightest, at $r\sim 35^\prime$, regions; their 
estimation $kT\lesssim 0.17$\,keV for the W\,50 shell from the non-detection implies 
an age of W\,50 of $\gtrsim 50000$\,yr as determined from the Sedov solution for 
a SNR. Things 
have changed with the better performances of the XMM-Newton observations 
of the eastern region of W\,50 (\cite[2007]{Bri07}; the following parameters 
are only for the eastern X-ray lobe): in the bright knot at $r\sim 35^\prime$ 
and in the sampled region closer to SS\,433, the temperature turns out rather 
$\approx 0.22$\,keV, ten times smaller, and the filamentary X-ray emission in 
the radio ear, supposedly of the jet 
terminal shock, has a temperature of $\approx 0.28$\,keV; besides, spectra of 
these regions are consistent roughly with the solar abundances, have a hard 
power law component, and all W\,50 in the field of view of the observations 
shines faintly in the soft X-ray, $<2$\,keV. Luminosity, $\gtrsim 6$--$8\cdot 
10^{34}$\,erg/s for the power law plus thermal components, and number density,
 $n_\mathrm{e} \sim 0.2$--0.7\,cm$^{-3}$, were estimated only for the bright 
knot. Unfortunately, the detail distribution of physical parameters over the lobe 
was not determined, and those parameters that are found, are not
proper parameters of the jet surroundings, being extracted from the radiation 
projected over different regions of W\,50.

The eastern X-ray lobe has the very contrasty narrow conical core of hard 
emission, $>1$\,keV, mainly of the power law nature (\cite[2007, Fig.~8]{Bri07}). 
This core is more or less uniform, has a flat transverse profile, its opening is 
$\sim 30^{\circ}$, although it subtends an angle of $18^{\circ}$ at SS\,433. The 
apex of the hard emission core falls slightly above the precession axis, at a 
distance from SS\,433 a little larger than $15^\prime$. It is seen also 
on Fig.~\ref{Xcone}, where the most contrasty image of the lobe, in the 
band 1.0--2.0\,keV, is overlaid on a radio image of W\,50. 
At the energies $<1$\,keV, this core is not seen against a diffuse background
(\cite[2007, Fig.~8]{Bri07}), though a hint of the separation between the core and
the surroundings could be in the two small troughs in the cut of the 
ROSAT image in the band $<1$\,keV (\cite[1996, Fig.~8]{Bri96}): they are 
separated by $18^{\circ}$ in azimuth relatively SS\,433, nearly symmetrically 
relatively the maximum in the cuts at higher energies. Such X-ray core is thought 
to map the hollow conical jet per se, which is recollimated and separates an 
interior spine from an exterior cocoon, and thus the hard emission of the lobe 
is localized to the jet and its neighborhood. The hollow morphology of the 
jet is supported also by the ring-like X-ray structure in the ear (\cite[2007]{Bri07}).
The more wide spread soft emission of the lobe increases barely to the jet axis. 
We accept therefore that this soft emission has a thermal component, as argued 
by \cite[(2007)]{Bri07} for the different regions 
of the lobe. The indistinguishability of the spin at energies $<1$\,keV means that 
temperatures of the cocoon and spin are nearly the same, and equal the 
temperature $\approx 0.22$\,keV of the two region of \cite['s\ (2007)]{Bri07}
spectral analysis. 

Now, we are going to estimate density of the thermal gas of W\,50 in the jet 
surroundings. For this purpose we make use of the normalization of the thermal 
component of the knot flux, found by \cite[(2007)]{Bri07} (actually an intricate 
problem even for the most bright knot) -- an unabsorbed flux of $F_\mathrm{0} 
= 9.9 \cdot 10^{-12}$\,erg/cm$^2$\,s in the energy band 0.2--2\,keV at a 
distance of 5\,kpc, -- and scale $F_\mathrm{0}$ to the thermal flux of the core
using the radial profile of the lobe brightness in the 
0.5--4.5\,keV energy band, obtained by \cite[(1983, Fig.~2)]{Wat83} from the cut 
over the azimuth range $2\theta_\mathrm{r} = 16^{\circ}$, indicated in 
Fig.~\ref{Xcone} (note that this cut overlays the hard emission core not 
exactly). From this profile, a ratio of the absorbed count fluxes of the core, taken 
between the distances $15^\prime$ and $33^\prime$, and the knot, between 
$33^\prime$ and $42^\prime$, is $a \equiv f_\mathrm{c}'/ f_\mathrm{k}' =0.62$, 
where the indexes c and k mean the core and knot, and the prime means 
an absorbed value. We note that $f'$ is the background subtracted count flux in the 
0.5--4.5\,keV energy band, integrated over an area of the cut. The hardness 
ratio $f'(>1.5\,\mathrm{keV}) /f'(<1.5\,\mathrm{keV})$, over the ROSAT energy 
range 0.1--2.4\,keV, is nearly constant through the core and knot
(\cite[1996, Fig.~9]{Bri96}). Besides, the latter two are similar in a spectrum 
(\cite[2007]{Bri07}). Hence, they are nearly equal in a ratio $h \equiv 
f_\mathrm{pl}/ f_\mathrm{th}$ of fluxes of the power law and thermal components, 
i.e. $h_\mathrm{c} \approx h_\mathrm{k}$. Then a ratio of the unabsorbed count 
fluxes of the core and knot equals approximately one of the absorbed fluxes: 
$f_\mathrm{c}/ f_\mathrm{k} \approx a$. With substitution of the decomposition
$f_\mathrm{c} = f_\mathrm{c\,th} + f_\mathrm{c\,pl} =  f_\mathrm{c\,th}
(1+h_\mathrm{c})$, this equation reduces to the proportion
\beq
a \approx \frac{f_\mathrm{c}} {f_\mathrm{k}} \approx \frac{f_\mathrm{c\,th}} 
{f_\mathrm{k\,th}} = \frac{F_\mathrm{c\,th}}{F_\mathrm{0}}
\label{c_th}
\eeq
where one accounts for the parity of the core and knot in $h$ and temperature.
What else should be do is to translate the 0.2--2\,keV energy flux 
$F_\mathrm{c\,th}$, emitted by the chosen core region, into the total thermal 
flux. The translation coefficient of $\approx 3.1$ was determined from the APEC model 
spectrum (V. Doroshenko, private communication) of an optically-thin coronal 
plasma of the solar abundances, and of temperature 0.22\,keV. Upon these 
the corresponding total thermal luminosity of the core is $L_\mathrm{c\,th} 
\approx 3.8 \cdot 10^{34}$\,erg/s $D_{4.5}^2$, where $D_{4.5}\equiv D/4.5$\,kpc. 

The region of the clearly discerned soft emission is symmetrical about the jet 
axis and subtends an angle of $2\theta_\mathrm{s}\sim 60^{\circ}$ at SS\,433 
(\cite[1983, Fig.~3]{Wat83}; \cite[2007, Fig.~8]{Bri07}). A volume of the segment 
of this region, which contributes into the radial profile of \cite[(1983)]{Wat83} 
between distances $r_1 = 15^\prime$ and $r_2 = 33^\prime$, can be approximated 
by the formula for a cone truncated on the top and the two opposite sides,
neglecting the inclination of the jet axis to the plane of the sky, as follows:
\beq
V_\mathrm{c} = \frac{\pi-2\phi+\sin{2\phi}} {3} \tan^2{\theta_\mathrm{s}} 
(r_2^3-r_1^3) = 4.1\cdot 10^{58}\,\mathrm{cm}^3,
\label{Vol}
\eeq
where $\phi=\arccos(\tan{\theta_\mathrm{r}}/\tan{\theta_\mathrm{s}})$.
A homogeneous number density of the jet surroundings in the eastern X-ray lobe
we estimate as $n_\mathrm{e} = \sqrt{\chi L_\mathrm{c\,th}/\Lambda(T) 
V_\mathrm{c}} \approx 0.15$\,cm$^{-3} D_{4.5}^{-1/2}$. Here, the equilibrium 
cooling rate $\Lambda = 10^{-22.32}$\,erg cm$^3/$s from (\cite[1993]{Su93}) 
and the electron to ion number density ratio $\chi \equiv n_\mathrm{e}/
n_\mathrm{i}=1.1$ for the solar abundances are accepted. For a consistency 
check, our estimation of the knot density is $n_\mathrm{e} \approx 
0.56$\,cm$^{-3}$ vs. \cite['s\ (2007)]{Bri07} 0.2--0.7\,cm$^{-3}$. Fortuitously,
nearly the same estimation have been given by \cite[(1983)]{Wat83} for the lobe 
in whole, $n_\mathrm{e} \sim 0.1$\,cm$^{-3}$ for the distance 5.5\,kpc, 
although they used the ten times larger temperature. The ready density estimation 
$n_\mathrm{e} \propto \sqrt{L_\mathrm{c\,th}/V_\mathrm{c}}$ gives only an upper 
limit of the average density in a heterogeneous medium, a case anticipated in the
jet vicinity. Here, we used the miniscule information contained in the X-ray 
observational data. In principle, an exceptional geometry of the X-ray lobe -- 
axisymmetric one, with the axis lying closely to the plane of the 
sky -- allows to restore tridimensional axisymmetric distribution of the lobe 
emissivity and, therefore, of the gas density. 

The temperature and density of the X-ray lobe bound tightly those of the W\,50
interior, which we consider here only beyond the first $\sim15^\prime$ from the
center. W\,50 
is rather of the same density as the lobe, $n_\mathrm{e} \sim 0.1$\,cm$^{-3}$: 
the jet ejected mass is nothing but $2 L_\mathrm{k0} t_\mathrm{j}/ 
v_\mathrm{j0}^2 \approx 5\cdot 10^{-3}$\,M$_\odot$ over the tentative lifetime 
of $t_\mathrm{j} = 10^4$\,yr. And W\,50 would have a temperature of 
$\sim 0.1$\,keV to be yet observed in whole in the soft X-ray emission
(\cite[2007]{Bri07}). It can be seen also on the particular cut of W\,50, 
shown by \cite[(1996, Fig.~8)]{Bri96}, that in the energy band $<1$\,keV, 
dominated by the thermal emission, the X-ray lobe is hardly discernible against 
W\,50, while the latter is noticeable against the background. Then pressure 
within W\,50 is $P_\mathrm{W50} = (1+1/\chi) n_\mathrm{e} kT \sim 3 \cdot 
10^{-11}$\,erg/cm$^3$. Its increase only several times will make W\,50 overall
as bright as the X-ray lobes. Surprisingly, this is nearly the pressure claimed by 
\cite[(1983)]{Eic83} for the recollimation of SS\,433's jets. We note that a 
non-thermal pressure in W\,50 of $0.56\cdot 10^{-11}$\,erg/cm$^3$ is 
significantly smaller than $P_\mathrm{W50}$, by our estimation from the data of 
\cite[(1998)]{Dub98} on the synchrotron radio emission of the 
spherical component; and the thermal pressure found in the X-ray lobe is inconsistent
with the gas density $\sim 0.01$--1\,cm$^{-3}$ in the eastern optical filaments,
reported by \cite[(2010)]{Abo10}.

\section{\large Origin of the radio nebula W\,50}
\label{Age}
Given the W\,50 inner pressure $P_\mathrm{W50}$, we could elucidate the age 
of W\,50, its origin and the role of SS\,433's jets in the nebula evolution. The 
spherical component of the nebula W\,50 could be inflated by a supernova or 
a powerful wind. In both cases, while the bubble has passed the first short stage, 
of free expansion into the interstellar medium, the radius of its forward shock
develops for long as
\beq
R(t) = \xi (E_0/\rho_0)^{1/5} t^b,
\label{SSS}
\eeq
where $\xi= 1.15$ and 0.88 for the SNR and wind, respectively, $\rho_0$ the 
density of the interstellar medium, which is assumed henceforth to be 
homogeneous, $b = 2/5$, $E_0$ the initial energy of the SNR, and 
$E_0 = L_\mathrm{w} t$ in the case of the wind with kinetic luminosity 
$L_\mathrm{w}$ (see \cite[(1977)]{Wea77} on a wind bubble, and e.g. 
\cite[(2015)]{Kim15} on a SNR). This is so called adiabatic phase, or the
Sedov phase in the case of a SNR. Further, after the formation of a dense, thin 
and cold shell behind the forward shock, as a result of radiative cooling of the 
swept-up and compressed interstellar gas, the bubble expansion changes: in the 
case of a wind bubble, only in constant $\xi$, 0.76, and a SNR obeys a new law, 
$R\propto t^{2/7}$, as is prescribed by the snowplow model of the shell, driven 
by pressure of the bubble hot interior through the interstellar medium.

W\,50 has likely passed the adiabatic phase, as it is evidenced by a faint optical
emission on the periphery of W\,50 (\cite[2007]{Bou07}). The interior of so mature 
bubble is suggested to be almost isobaric and to have nearly flat profiles of 
temperature and density due to thermal conduction (\cite[1977]{Wea77}; 
\cite[1992]{Cui92}), not to mention the possible turbulent mixing, driven by the 
jets. Moreover, in any scenario of the W\,50 origin, the powerful hypersonic 
wind from the super-Eddington accretion disc in SS\,433 system, of mass flow rate 
$\dot{M}_\mathrm{w} \sim 0.6 \cdot10^{-4} D_{4.5}^{3/2}$~M\sun/yr 
(\cite[2006]{Fuc06}; see also \cite[2009]{Per09}), would extend only to an inner 
shock in the bubble interior, before which the gas is cool, $\sim 2\cdot 10^4$\,K, 
and tenuous, $n_\mathrm{w} \propto r^{-2}$, and beyond the shock the gas 
becomes hot and dense (\cite[1977]{Wea77}). The shock radius would be of the 
order of $\sqrt{v_\mathrm{w} \dot{M}_\mathrm{w}/4\pi P_\mathrm{W50}} 
\sim 13$\,pc, or $\sim 10^\prime$, as determined from continuity of momentum 
flux at the shock jump, where $v_\mathrm{w} =1500$~km/s is the wind maximal 
velocity (e.g. \cite[2009]{Per09}). However, we should not rely on the isotropy 
assumed in this
estimation. Such inner structure of W\,50 conforms with the nearly flat appearance 
of the X-ray lobe, and with the ignition of the X-ray lobe at the radius  
$\sim 15^\prime$, supposedly the wind inner shock radius (\cite[1983]{Kon83}).

%******************************************************************
%
\begin{figure}[t]
\centering
\includegraphics[width=8cm]{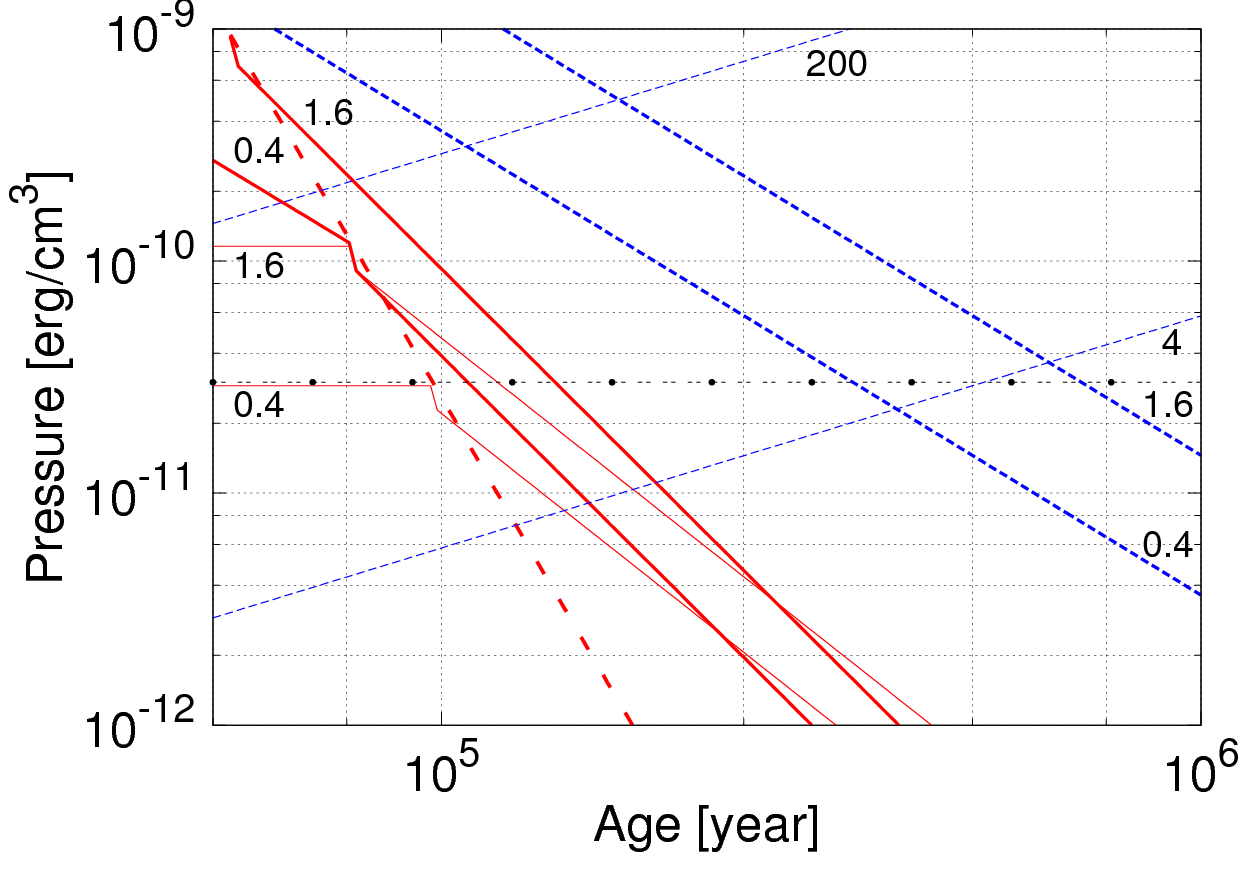}
\caption{
Contour map of the parameters of interstellar bubbles on the plane (age, 
pressure). The red solid and long dashed lines on the left -- for SNRs. 
The blue short dashed lines -- for wind blown bubbles. The red 
long dashed line -- the line of the shell formation in SNRs: the Sedov phase 
(adiabatic) lies on the left of this line. The thick lines -- the interstellar 
hydrogen density contours, in units of cm$^{-3}$. The thin lines -- the SNR 
initial energy contours, in units of $10^{51}$\,erg, and the wind power 
contours, in units of $10^{37}$\,erg/s. The line of circles -- the observed 
pressure of the W\,50 nebula.
}
\label{age}
\end{figure}
%
%******************************************************************

A mean pressure of a bubble is determined as $P= (\Gamma-1) E_\mathrm{th}/V$,
where $\Gamma = 5/3$ is the adiabatic index for thermal plasma, $E_\mathrm{th}$
the thermal energy of the bubble interior of volume $V = 4\pi R^3/3$. In
the Sedov phase (the adiabatic phase), this energy is constant, 
$E_\mathrm{th} = E_\mathrm{0\,th} \equiv 0.717E_0$, and in the snowplow 
phase the SNR energy decreases due to adiabatic and radiative losses as 
$E_\mathrm{th} = 0.8 E_\mathrm{0\,th} (r_\mathrm{sf}/R)^2 
(t_\mathrm{sf}/t)$, the function fitted by \cite[(2015)]{Kim15} to their 
numerical simulations. Here 
\beq
t_\mathrm{sf} = 4.0\cdot 10^4 E_{51}^{0.22} n_0^{-0.59}\,\mathrm{yr},\
r_\mathrm{sf} = 22.1 E_{51}^{0.29} n_0^{-0.43}\,\mathrm{pc}
\label{shell}
\eeq
are the time and radius at the shell formation epoch,
$E_{51} \equiv E_0/10^{51}$\,erg, $n_0 \equiv n_\mathrm{H}/1$\,cm$^{-3}$,
and $n_\mathrm{H}$ the interstellar hydrogen density. In the case of a wind 
bubble, $E_\mathrm{th}= (5/11) E_0$ in both the adiabatic and the idealized snowplow 
phases, that however doesn't account for the radiative cooling of the hot interior 
gas, which can be significant at large times, $>100\,000$\,yr in the particular 
case of $L_\mathrm{w}  \sim 10^{36}$erg/s (\cite[1977]{Wea77}). Using the above 
formulated laws $R(t)$ and $E_\mathrm{th}(R,t)$, we have plotted the contours of $E_0$,
$\rho_0$ and $L_\mathrm{w}$ on the plane ($t,\ P$) in Fig.~\ref{age}, where $P$
is the mean pressure inside a spherical bubble of the radius 38\,pc 
of W\,50's spherical component (at the distance 4.5\,kpc). The interstellar density 
contours for both models are $n_\mathrm{H} = 0.4$ and 1.6\,cm$^{-3}$, 
that overlaps the initial density range of the 
HI gas in W\,50's neighbourhood, found by \cite[(2007)]{Loc07} and scaled to 
the distance 4.5\,kpc. The SNR energy contours $E_0 = 0.4$ and 1.6 times 
$10^{51}$\,erg just overlap the typical energy $10^{51}$\,erg of SNRs. The wind 
power contours $L_\mathrm{w} = 4\cdot 10^{37}$ and $2\cdot 10^{39}$\,erg/s 
demarcate a region between the kinetic luminosities of the supercritical accretion 
disk wind and of the hypothetical wind of the jets: as if they energize W\,50 
isotropically.

In view of the observed pressure $P_\mathrm{W50} \sim 3 \cdot 
10^{-11}$\,erg/cm$^3$, the line of shell formation in Fig.~\ref{age}, determined 
by Eqs.~(\ref{shell}), restricts the age of W\,50 as a SNR by $t_\mathrm{W50} 
\ge 97\,000$\,yr, and an intersection of the admitted $n_\mathrm{H}$ and $E_0$ ranges 
corresponds to the ages just over $100\,000$\,yr -- cf. the most probable lifetimes 
10\,000--100\,000\,yr of SS\,433 as a binary system undergoing a super-Eddington
accretion, estimated by \cite[(2000)]{Kin00}. Further, the disk wind scenario 
matches well the observed pressure, but at the cost of the uncomfortable large ages 
$\sim 300\,000$--700\,000\,yr. However, the unaccounted radiative cooling of the hot
interior gas of a wind bubble would shift the bottom parts of the wind contours 
to the lower ages in the same manner as it bends the SNR contours while they pass
the shell formation line. With respect to the jets
as a source of the wind, without a preexisted SNR, they would overwhelm 
W\,50 in the energy in $\sim 10$ times during a lifetime of $> 100\,000$\,yr.
So, the wind scenario with the admitted $n_\mathrm{H}$ range issues in either 
the improbable large age or exaggerated pressure. This is due to the fact that 
a continuous injection of energy is less effective way to blow a bubble than 
an initial blast, as it is seen from the difference between both models in the 
constant $\xi$ in Eq.~(\ref{SSS}). The exaggeration of pressure seems to be 
proper to the mix scenario as well, when the jets are fully slowed down in the 
interior of a preexisted bubble. In reality, the decelerated jets would distort 
sphericity of the bubble. All this suggests that the radio nebula W\,50 was 
initiated by a supernova explosion well near $100\,000$\,yr ago, and the 
SS\,433 jets propagate in the preexisted SNR and have to deposit most of their 
huge kinetic energy beyond the SNR radius, unless their lifetime is small enough:
\beq
t_\mathrm{j} \ll E_0/2\zeta L_\mathrm{k0} \sim 16\,000 (E_{51}/\zeta)\,\mathrm{yr}
\label{tjet}
\eeq
where $\zeta$ is the ratio of the jets energy transfered to the SNR to the total 
energy of the jets. This SNR would have an initial energy in the range 1.0--3.7$\cdot 
10^{51}$\,erg, bounded by the interstellar density range $n_\mathrm{H} =
0.4$--1.6\,cm$^{-3}$.

The above picture on the origin of W\,50 is built on the assumption that the 
interstellar medium is homogeneous. In the case of the inhomogeneity, the bubble 
expands faster, therefore its age for a given radius is smaller, that
depends on the volume filling (e.g. \cite[2015,\ Fig.\,4]{Kor15}). At that, the 
relative roles of the SNR, disk wind and jets in the origin of W\,50, discussed 
above, are thought not to change. We note that just the inhomogeneity
allows so small age as $\sim 100\,000$\,yr in the wind blown bubble model
devised by \cite[(1983)]{Kon83}.

\section{Deceleration of the SS\,433 jets in the X-ray lobes}
\label{Entr}
To link the jets of SS\,433 observed downstream of the source to only $\sim 6''$ 
with the jet signatures 
at parsec scales is a challenging problem. In the innermost regions of W\,50, 
the jet travels through the low density cavity, evacuated by the hypersonic wind 
from SS\,433. Just before the wind shock, at the suggested radius $r_1 \sim 15^\prime$,
the isotropic wind hydrogen density $n_\mathrm{w}(r_1) = \dot{M}_\mathrm{w} / 
1.4 m_\mathrm{H} 4\pi r_1^2 v_\mathrm{w} \sim 3\cdot 10^{-4}$\,cm$^{-3}$ 
is much smaller than the observed density of the X-ray lobe, $m_\mathrm{H}$ 
being the mass of hydrogen. Going through 
this shock into the significantly more denser medium, some hundreds 
times, the jet becomes strongly underpressured, and possibly realigns to the state
of pressure equilibrium with the surroundings. There are not seen separate strong 
shock waves, but abrupt appearance of the X-ray lobe. The conspicuous conical 
core of the lobe, of half-angle $\theta_\mathrm{c} \sim 15^{\circ}$, suggests that 
the jet is recollimated to the core opening. Moreover, the jet  should be hollow: if all 
mass within the cone, of mass density $\rho_\mathrm{a} = (1+1/\chi)\bar{\mu} 
n_\mathrm{e}$, where $n_\mathrm{e}\sim 0.1$\,cm$^{-3}$ and $\bar{\mu} = 0.62 
m_\mathrm{H}$, were involved into the jet, the later would be as slow as 
\beq
\frac{1}{r_1 \tan \theta_\mathrm{c}}
\sqrt{\frac{\Pi_\mathrm{j}}{\pi \rho_\mathrm{a}}} \sim 700\,\mathrm{km/s}, 
\label{vel}
\eeq
at the X-ray lobe beginning due to conservation of the jet momentum flux 
$\Pi_\mathrm{j}$, that is $\sim$ a hundredth of the initial velocity, and means
almost full dissipation of the jet kinetic energy. This does not stand out in 
brightness and, therefore, is improbable; besides, the jet should yet thrust the ear 
and provide the ring like morphology of the terminal shock at distances 
$r\sim 60^\prime$ (\cite[2007]{Bri07}). Hereafter, we accept this geometry of 
the jet, with the tentative half-angle $\theta_\mathrm{j} \sim 4^{\circ}$ of the 
filled surface layer of the hollow conical jet, that is ready by the jet nutation 
at the origin (\cite[1987]{Bor87}), and
can be substantiated by width of the brightness depression at the boundary of 
the hard X-ray core, observed in the soft X-ray cut, that was discussed in 
Sect.~\ref{PhC}. 

The fact of observation of the more or less smooth X-ray lobe suggests 
continuous deceleration of the jet of SS\,433. In general, a jet spends momentum via
viscous stresses at the interface with the ambient medium. The viscosity probably is 
dominated by turbulence and magnetic fields. Indeed, in two shearing flows, the boundary
between them is inevitably unstable, that gives rise to turbulent mixing of the fluid 
in the boundary layer (\cite[1987, \S\S 29, 36]{LL87}). Thereafter, this layer 
ingests mass from both flows: between vortex-free and vortex layers fluid flows 
only from the first to the second (\cite[1987, \S 35]{LL87}). As a result, in the 
case of a jet, the latter entrains the ambient medium. This picture is supported 
by the observed evolution of a transverse velocity profile of the gradually decelerating 
jets of Fanaroff-Riley class I (FR\,I) radio galaxies, and by a conservation-law 
analysis of these jets (see \cite[(2014)]{Lai14} for a comprehensive study of 
the kinematics and dynamics of FR\,I jets). However, deceleration of 
relativistic jets is not predictable by existing models. We note that the jets 
of SS\,433 similarly to FR\,I jets have not a prominent hot spot at the end, 
contrary to FR class II jets, which are essentially relativistic until the hot spot. 

\subsection{Mechanism of mass entrainment in relativistic jets}
\label{Entr1}
\begin{figure}[t]
\centering
\includegraphics[width=8cm]{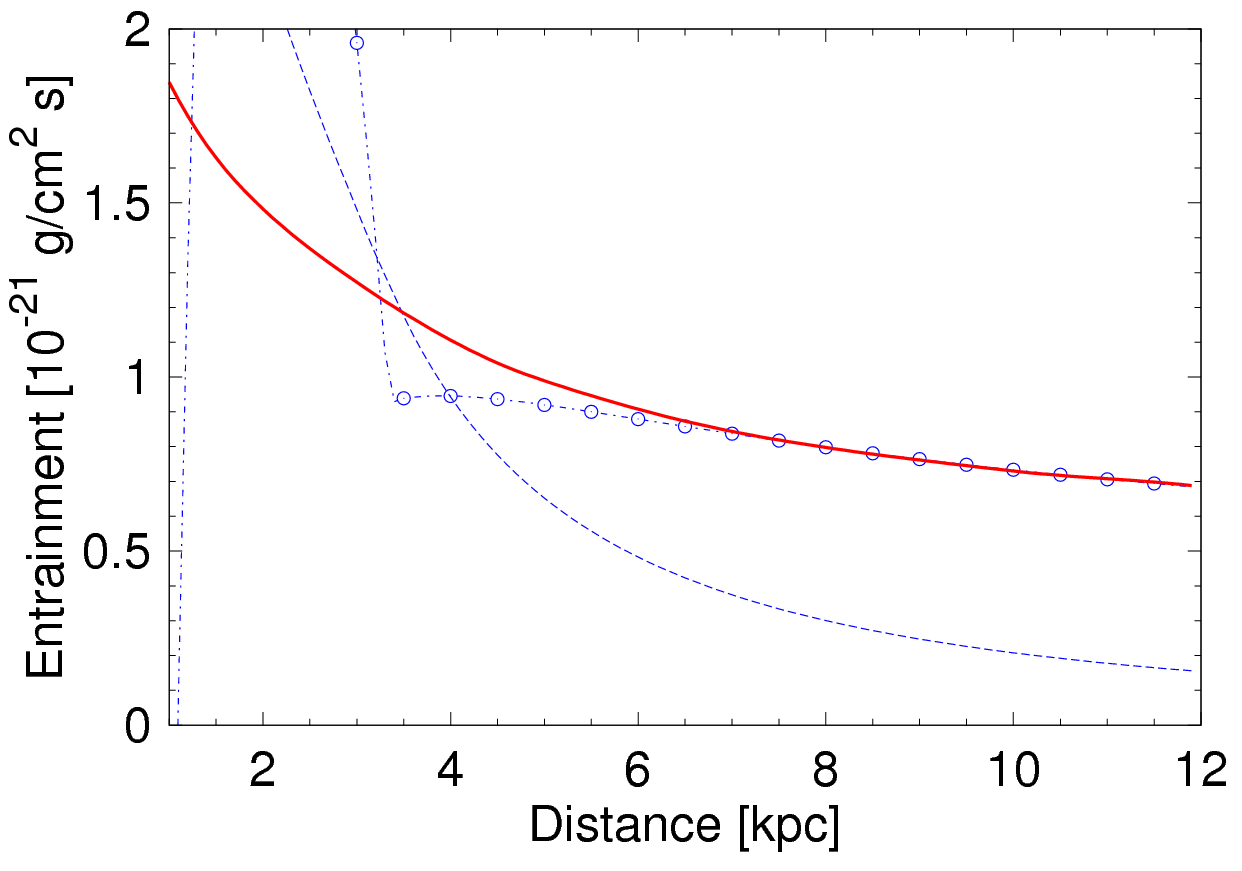}
\caption{
The entrainment along 3C\,31's jets: the flux densities of the external 
entrainment (blue dash-dotted line with circles) and of the internal entrainment  
(blue dashed line) after \cite[(2009)]{Wan09}, and the 
theoretical external entrainment of the form (\ref{deer}) (red solid line).
}
\label{3C31}
\end{figure}

For the jets of the FR\,I radio galaxy 3C\,31, \cite[(2002)]{Lai02} derived 
the semiempirical function of the rate of mass injection into the jet (or, 
rather, the entrainment rate) in dependence on the jet distance 
$r$ from the galaxy nucleus on the basis of the laws of conservation of mass, 
energy and momentum. Further, \cite[(2009)]{Wan09} derived the entrainment 
rate for the case of equilibrium in pressure of these jets with their surroundings.
Fig.~\ref{3C31} shows the longitudinal profile of mass flux $\dot q_\mathrm{e}$ 
of the entrained surrounding matter per unit area of the jet surface, henceforth 
the external entrainment, for the 3C\,31 jets, obtained from 
\cite['s (2009)]{Wan09} boundary-layer entrainment $g$ and internal entrainment
$g_\mathrm{s}$, which could be from the supposed jet-contained stars, as follows:
\beq
\dot q_\mathrm{e} = \frac{1}{2\pi R_\mathrm{j}}
\frac{\mathrm{d}}{\mathrm{d}r}(g - g_\mathrm{s}), 
\label{def_q}
\eeq
where $R_\mathrm{j}$ is the jet radius, given by the jet geometry from 
(\cite[2009]{Wan09}). However, it is not clear how much correct is the 
external entrainment, defined in this way, just subtracting the internal 
entrainment, because \cite[(2009)]{Wan09} didn't account for the internal 
entrainment in their model of the boundary-layer entrainment, while the latter 
is the total entrainment required by the jet kinematics.
The internal entrainment rate normalized to unit area of the jet surface is also 
shown in Fig.~\ref{3C31}. As \cite[(2009)]{Wan09} have noted, uncertainties of 
the internal entrainment defined by their Eq.~(37) are large. 

%
%******************************************************************
%
%
%******************************************************************

Looking for a formalism of the entrainment, which would hopefully fit any 
relativistic jet, we have confronted the profile $\dot q_\mathrm{e}(r)$ in 
Fig.~\ref{3C31} with a simplistic model. Let in a frame comoving with the mean 
local flow in the medium adjacent to a jet, the velocity of turbulent irregular 
pulsations in this flow is $v_\mathrm{t}$, and the velocity of the jet 
$v_\mathrm{j}$. The pulsating component of a movement we restrict for 
the sake of simplification only to six reciprocally orthogonal directions, e.g. 
along the Cartesian coordinate axes, with $X$ axis
aligned with the vector ${\bf v}_\mathrm{j}$, so that in the comoving 
frame at any point of the average jet boundary turbulent fluid moves into the 
jet only one sixth of time. Hence, the density of mass flux into the jet is
$\rho_\mathrm{a} v_\mathrm{t}/6$, $\rho_\mathrm{a}$ being the mass density 
of the ambient medium. A part of the mass flux is swallowed by the jet, and 
the remainder continues the turbulent dancing with the adjacent flow. Namely, 
we impose that the jet absorbs only the component of momentum $p$ of the 
turbulence pulse in the jet rest frame, which is perpendicular to the jet 
boundary, i.e. a proportion of the absorbed part of the momentum is 
\beq
\sigma_\mathrm{e} \equiv \frac{p_\perp}{p} = \frac{v_\perp}
{\sqrt{(v_\parallel^2+v_\perp^2)}} = \frac{\eta}{\sqrt{\gamma^2+\eta^2}}, 
\label{saber}
\eeq
under constancy of mass density of the turbulent fluid, where $v_\perp =
v_\mathrm{t}/\gamma$ is the perpendicular component of the velocity of the 
turbulence pulse in the jet rest frame, $v_\parallel = v_\mathrm{j}$ the 
parallel component, $\eta= v_\mathrm{t}/v_\mathrm{j}$, and  
$\gamma = (1-(v_\mathrm{j}/c)^2)^{-1/2}$ is the Lorentz factor (for relativistic
transformations of velocity see e.g. \cite[1971, \S 5]{LL71}). 
The parameter $\sigma_\mathrm{e}$ is in essence the cross-section of the 
entrainment process. Note that $\sigma_\mathrm{e} = \sin i$, where $i$ is the 
inclination of the velocity of the turbulence pulse to the jet boundary in the jet 
rest frame. Then, the flux density of the entrainment is
\beq
\dot q_\mathrm{e} = \sigma_\mathrm{e} \beta \rho_\mathrm{a} v_\mathrm{t} =
\alpha \frac{P}{v_\mathrm{j}}, 
\label{deer}
\eeq
where $\beta = 1/6$, $P= \rho_\mathrm{a} c_\mathrm{s}^2/\Gamma$ 
the external pressure, $c_\mathrm{s}$ being the sound speed in the external 
medium, 
\beq
\alpha = \beta \Gamma M_\mathrm{t}^2 \frac{1}{\sqrt{\gamma^2+\eta^2}} 
\label{alpha}
\eeq
the dimensionless viscosity parameter, and $M_\mathrm{t}= v_\mathrm{t}/c_\mathrm{s}$
the Mach number of the turbulence. The matter entrained from the surroundings 
is mixing with the jet matter, and accelerating to the jet velocity $v_\mathrm{j}$, 
that should be provided by the flux of the jet momentum
\beq
\sigma = \dot q_\mathrm{e} v_\mathrm{j} = \alpha P 
\label{str}
\eeq
per unit area of the jet surface, transfered normally to the jet boundary. That 
is just definition of a viscous stress. It appears that the viscous stress 
(\ref{str}) at the jet boundary has the functional form just as is in 
\cite['s (1973)]{Sha73} $\alpha$-model for accretion disks, and as have been 
derived by \cite[(1982)]{Beg82} for extragalactic jets to explain their heating 
via viscous stresses.

To plot the entrainment function~(\ref{deer}), we used the profiles of pressure
$P(r)$, density $\rho_\mathrm{a}(r)$ and velocity $v_\mathrm{j}(r)$ from
(\cite[2009]{Wan09}). A Mach number of the turbulence of $M_\mathrm{t}=1$
was accepted as the best guess for supersonic jets. In Fig.~\ref{3C31}, this 
function fits the overall data on the 3C\,31 jet, the magnitude and slope of the 
external entrainment profile, and does it excellently for the jet region beyond 
$\sim 6$\,kpc if $1/5$ is used instead of the coefficient $\beta = 1/6$ in 
Eq.~(\ref{deer}). The worse match before $\sim 6$\,kpc can be explained 
partly by a difference between the flaring region at 1.1--3.5\,kpc, 
where the jet realigns (\cite[2009]{Wan09}), and the more quiescent 
outer jet, and, on the other hand, by the 
increase of the internal entrainment upstream of the jet (Fig.~\ref{3C31}), 
which is crudely estimated and subtracted to get the external entrainment. 
Such successful fit encourages us to accept the model and apply it to the SS\,433
jets, under the coefficient $\beta = 1/5$.

\subsection{Entrainment model of deceleration of the SS\,433 jets}
\label{SS}
How much the velocity of the SS\,433 jet changes with the distance due to the
injection of matter from the ambient medium we can derive straightforwardly from 
the relationship $v \propto \Pi/\dot M$ between momentum flux $\Pi$, which is
constant, and mass flux $\dot M$. The derivation is following. For the jet flow 
of the rest-frame mass density $\rho_\mathrm{j}$ and enthalpy 
$\omega = \rho_\mathrm{j} c^2 + \epsilon + P$, where $\epsilon$ is the 
internal energy density, and $P$ is the pressure, with velocity vector 
${\bf v}_\mathrm{j}$ and Lorentz factor $\gamma$, the flux densities 
of mass and x-component of momentum through the surface piece, which 
outward unit normal vector is $\bf n$, are respectively
\bea
\dot q = \gamma \rho_\mathrm{j} ({\bf nv}_\mathrm{j}),
\label{qfr}\\
\dot p_\mathrm{x} = \gamma^2 \frac{\omega}{c^2} v_\mathrm{jx} ({\bf nv}_\mathrm{j}) 
+ Pn_\mathrm{x} \approx \gamma \dot q v_\mathrm{jx} + Pn_\mathrm{x},
\label{pfr}
\eea
where index x means the component of a vector along the jet axis. The right part
of Eq.~(\ref{pfr}) is derived neglecting the terms $\epsilon$ and $P$ in the 
enthalpy $\omega$ of the mild relativistic jets of SS\,433.

%
%******************************************************************
%
\begin{figure}[t]
\centering
\includegraphics[width=8cm]{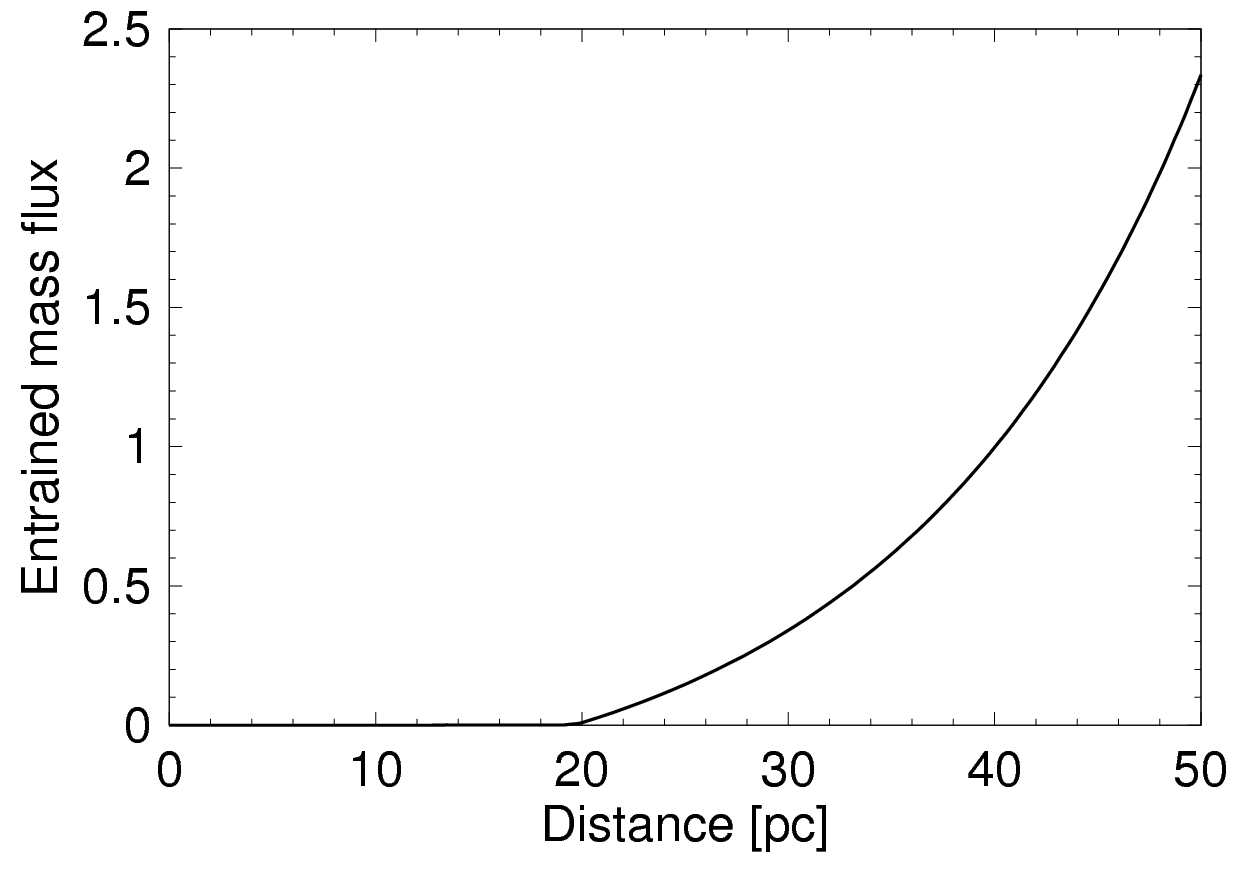}
\caption{
The mass flux along the SS\,443 jet provided by the entrainment, 
$\dot M_\mathrm{e}$, in units of initial mass flux, $\dot M_\mathrm{j0}$.
}
\label{Mflux}
\end{figure}
%
%******************************************************************

Let the closed surface of an integration of the flux densities consists of the 
two cuts of the jet by the spheres of radii $r_0$ and $r$, 
centered at the jet source, and the jet boundary surfaces between them, so that
the velocity vector ${\bf v}_\mathrm{j}$ is normal to the cuts and tangential to
the boundary surface. The jet density and velocity are 
supposedly functions of only the distance $r$, i.e. have flat profiles over the 
spherical cuts, that could be provided by turbulent mixing. Then, for the hollow
conical jet. the mass fluxes in absolute value are
\beq
\dot M_\mathrm{j}(r) = \gamma \rho_\mathrm{j} v_\mathrm{j}S,
\label{Mj}
\eeq
through the jet cut of an area $S$, and
\beq
\dot M_\mathrm{e} = 2\pi 
(\sin(\theta_\mathrm{c}-\theta_\mathrm{j})+\sin(\theta_\mathrm{c}+\theta_\mathrm{j}))
\int_{r_0}^r \dot q_\mathrm{e}(\zeta) \zeta d\zeta,
\label{Mentr}
\eeq
through the jet surfaces between the cuts, where the entrainment flux density is 
assumed the same on the both sides of the hollow jet. Over the chosen above 
closed surface, the integrations of the densities (\ref{deer}, \ref{qfr}) for 
the flux of mass, and (\ref{pfr}) for the flux of momentum give zeros: 
\bea
-\dot M_\mathrm{j}(r_0) - \dot M_\mathrm{e} + \dot M_\mathrm{j}(r)=0,
\label{Qfr}\\
-f(\gamma \dot M_\mathrm{j} v_\mathrm{j} + PS)_{r_0} - F + 
f(\gamma \dot M_\mathrm{j} v_\mathrm{j} + PS)_{r} =0,
\label{Pfr}
\eea
in the case of the jet in a 
steady state, one more idealization of the model. Here, the first and third parts 
in both equations are the fluxes through the spherical cuts at radii $r_0$ and $r$, 
respectively, $f$ the modulus $|n_\mathrm{x}|$ averaged over a cut, and the 
momentum injected through the jet boundary surface by the entrained matter is 
neglected, because it is miniscule. The buoyancy term
\beq
F = 2\pi (\sin(\theta_\mathrm{c}+\theta_\mathrm{j})-
\sin(\theta_\mathrm{c}-\theta_\mathrm{j})) \int_{r_0}^r P d\zeta
\label{buoy}
\eeq
summed with other members of Eq.~(\ref{Pfr}), containing pressure $P$, results 
in zero in the case of pressure equilibrium of the SS\,433 jet with the ambient 
medium, assumed to be isobaric. Then, Eq.~(\ref{Pfr}) is reducing to the 
anticipated ratio of the jet velocities at radii $r$ and $r_0$
\beq
\frac{v_\mathrm{j}(r)}{v_\mathrm{j}(r_0)} = \frac{\gamma(r_0)}{\gamma(r)}
\frac{\dot M_\mathrm{j}(r_0)}{\dot M_\mathrm{j}(r)} 
\approx \frac{\dot M_\mathrm{j}(r_0)}{\dot M_\mathrm{j}(r)},
\label{rvel}
\eeq
where the approximation by the right part has a maximal error of 3.6\%. 
Eqs.~(\ref{rvel}, \ref{Qfr}, \ref{Mentr}, \ref{deer}) have a solution
\beq
v_\mathrm{j}(r) = v_\mathrm{j}(r_0) \exp{\left(-\frac{\alpha A P 
\left(r^2-r_0^2\right)} {2\dot M_\mathrm{j}(r_0) v_\mathrm{j}(r_0)}\right)},
\label{velsol}
\eeq 
under the conditions $\eta \ll \gamma = 1$, and constancy of the viscosity 
parameter $\alpha$ and the pressure $P$, where $A=2\pi (\sin(\theta_\mathrm{c}-
\theta_\mathrm{j})+\sin(\theta_\mathrm{c}+\theta_\mathrm{j}))$.

The profile along the jet of the mass flux $\dot M_\mathrm{e}$ in Fig.~\ref{Mflux}, 
which follows from the system of Eqs.~(\ref{rvel}, \ref{Qfr}, \ref{Mentr}, 
\ref{deer}), shows a rate of the 
loading of mass from the surroundings into the SS\,433 jet as much as the initial 
jet mass flux $\dot M_\mathrm{j0} = 2L_\mathrm{j0}/v_\mathrm{j0}^2$. There 
were used the parameters $\beta = 1/5$ and $M_\mathrm{t} =1$, the jet geometry 
with $\theta_\mathrm{j} 
= 4^{\circ}$ and $\theta_\mathrm{c} = 15^{\circ}$, and the uniform 
physical conditions $P= 3 \cdot 10^{-11}$\,erg/cm$^3$ and $n_\mathrm{e} = 
0.1$\,cm$^{-3}$ in the region $r>20$\,pc (Sect.~\ref{PhC}). This loading issues
in a relative decrement of the jet speed of $\delta v_\mathrm{j} \equiv 1-
v_\mathrm{j}/v_\mathrm{j}(r_0) = \kappa/(1+\kappa) \approx 63\%$  
at the distance of the X-ray bright knot $\sim 35^\prime$ ($\sim 46$\,pc), 
i.e. yet before the entry into the ear, where $\kappa = 
\dot M_\mathrm{e}/ \dot M_\mathrm{j}(r_0)$, and the radius $r_0$ refers 
to the parameters at the jet beginning. The jet kinetic luminosity is sinking by
the same part, $1- L_\mathrm{k}/ L_\mathrm{k}(r_0)= \delta v_\mathrm{j}$.
The mass loading in the wind cavity region, at $r<20$\,pc, 
is insignificant, because of low pressure. In contrast, the density and pressure 
in the region of the shell, marked by the X-ray bright knot, 
should only rise, and the mass loading should do so. 

\section{Concluding remarks}
\label{Iss}
The morphology of the eastern X-ray lobe in W\,50 evidences a continuous 
proceeding of SS\,433's jet through the nebula on scales of dozens 
parsecs. At that, the jet is found to be recollimated from the opening 
$40^\circ$ to the $\sim 30^\circ$. The X-ray brightness distributions in the 
soft and hard energy bands of the XMM-Newton observations are consistent 
with the hollow structure of the jet. This picture suggests that the jet 
bypasses the X-ray brightest knot, which resides at the place of the shell
existed even before the jet.

From the X-ray observations presented in (\cite[1983]{Wat83}; \cite[1996]{Bri96}; 
\cite[2007]{Bri07}), we derived an electron density of $\sim 0.1$\,cm$^{-3}$ and 
a pressure of $\sim 3 \cdot 10^{-11}$\,erg/cm$^3$ for the thermal gas in the 
spherical component of 
W\,50. We note that only the XMM-Newton observations have allowed 
\cite[(2007)]{Bri07} to separate surely the thermal components of the X-ray emission 
of the bright knot and, less surely, of the region to the west of the knot.  
However, they have not characterized the physical conditions
in the X-ray lobe in whole. Solely a model of SNR dynamics predicts the 
observed pressure in conjunction with the reliable ages of SS\,433 
10\,000--100\,000\,yr (\cite[2000]{Kin00}), not a model of stellar wind bubble. 
Though, this is not conclusive in the case of an inhomogeneous ambient medium,
when the wind origin of W\,50 is possible too (\cite[1983]{Kon83}).
Such small observed pressure pinpoints W\,50 as an old SNR of age 
$\sim 100\,000$\,yr, which is obtained accounting for radiative losses of the 
SNR in the pressure driven phase, unlike all previous estimations of W\,50's age,
based on the Sedov model. In the case of an inhomogeneous
interstellar medium, the age would be smaller. An influence of 
the jets on the spherical component would be restricted to prevent a large 
distortion of the sphericity, therefore the jets exhaust most of their energy 
into the ears or are much younger than the SNR. The thermal energy content 
in the W\,50 nebula is 
$\sim (\Gamma-1) P_\mathrm{W50} (V_\mathrm{SNR} + V_\mathrm{E} 
+ V_\mathrm{W}) = 4.4\cdot 10^{50}$\,erg, where $V_\mathrm{SNR} =
6.6\cdot 10^{60}$\,cm$^3$, $V_\mathrm{E} = 2.4\cdot 10^{60}$\,cm$^3$,
and $V_\mathrm{W} = 6.9\cdot 10^{59}$\,cm$^3$ are the volumes of
the spherical component, the eastern and western ears, respectively.
Supposedly, the same amount of energy is contained in the kinetic energy
of the nebula's shell, then the total energy of W\,50 is $\sim 10^{51}$\,erg.

The simulations of \cite[(2011b)]{Go11b} provide an important view that 
the jet-SNR interaction doesn't influence the expansion of W\,50's spherical 
shell, and the jets are a relatively young phenomenon, $\lesssim 20\,000$\,yr,
the age they give to the nebula W\,50. The latter issues naturally from an 
amount of the jets momentum imparted to the ears, and hence from the ears 
size. Besides clarification of the role of the jets in the origin of W\,50, this 
justifies the application of theories of the spherical wind bubbles and SNRs to 
the evolution of W\,50. However, so small age of W\,50's spherical component, 
obtained in the simulations, contradicts not only to our result, based on 
a theory for the shock dynamics of SNRs, but also to the simulations on
SNRs elsewhere (e.g. \cite[2015]{Kim15}; \cite[2015]{Kor15}). 

\cite['s\ (2011b)]{Go11b} simulations invoke the intermittent activity of the 
jets for the observed shape of the ears. At that, the continuous activity,
coupled with the recollimation and deceleration, appeals more urgently  
as the observed non-unique properties of the broad class of jets, of radio 
galaxies of Fanaroff-Riley Class I (\cite[2014]{Lai14}). There seems the sameness 
between them and SS\,433's jets: the pressure jump in the surroundings 
(afforded possibly by the wind in the case of W\,50), beyond which the 
jets brighten, recollimate and decelerate (the latter two are obvious only for 
FR\,I jets), and the faintness of terminal shock at the jet head (contrary to 
the prominent hotspots of FR\,II jets) -- the characteristic of decelerated jets.
 
Indeed, the derived pressure of W\,50's interior turns out to be so large as to 
cause deceleration of the jet via the viscous jet-surroundings interaction. 
There is as yet no theory of this interaction for relativistic astrophysical jets. 
The studies of FR\,I jets suggest that these jets decelerate continuously 
via entrainment of the ambient medium. We devised the model of 
the entrainment, which successfully fits the semiempirical profile of 
entrainment rate for the jets of the radio galaxy 3C\,31, found by 
\cite[(2009)]{Wan09}. By this model, the entrainment results from turbulence 
at the jet surface, that exerts in effect a viscous tension in the 
jet $\sigma = \alpha P$, of the well known form in the theory 
of accretion disks, where the viscosity parameter $\alpha$ is defined by the model. 
This model predicts that the eastern jet of SS\,433 decelerates by $\sim 60$\%
in bounds of the circular shell of W\,50. 

Thus, the decelerated jet would inject $\sim$ a half of its energy into W\,50's spherical
component. For the sphericity, the age of the jets, Eq.\,(\ref{tjet}), should be 
much smaller than 27\,000\,yr, that is consistent with the age delimitation in 
(\cite[2011b]{Go11b}). Also, there seems to be a balance between the jet
power and the luminosity of the X-ray lobe. During tentative lifetime 
$t_\mathrm{j} \sim 10^4$\,yr the jet has put approximately a half the ejected energy 
$W_\mathrm{j} = L_\mathrm{k0} t_\mathrm{j} \sim 3\cdot 10^{50}$\,erg 
in the expansion of W\,50, another part has been deposited into
the thermal energy of the interior gas. While an estimation
$W_\mathrm{th} = L_\mathrm{th} t_\mathrm{th} \sim 3\cdot 
10^{49}$\,erg, where $L_\mathrm{th} \sim 10^{35}$\,erg/s is the thermal 
luminosity of the X-ray lobe (Sect.~\ref{PhC}), and $t_\mathrm{th} \sim 
kT/n_\mathrm{e} \Lambda \sim 10^7$\,yr the time of the radiative cooling of 
the lobe, gives for the thermal energy $\sim 1/10$ of the $W_\mathrm{j}$ vs. 
the above proportion $\sim 1/2$. The discrepancy by a factor of 5 is possibly
attributable to roughness of our estimations.

Further studies of the distribution of physical parameters in the nebula W\,50,
in particular on the basis of the available X-ray data, would improve the above 
picture of the evolution of SS\,433's jets on scales of dozens parsecs.

\section{Acknowledgements} We would like to thank Victor Doroshenko, Vladislav 
Stolyarov and Valery Suleimanov for a versatile invaluable help in the work on
the paper.

\end{document}